\documentclass[noinfoline,stslayout]{imsart}

\usepackage{amssymb}
\usepackage{amsmath}
\usepackage{graphicx}
\usepackage[sort,comma,numbers]{natbib}

\usepackage[mathletters]{ucs}
\usepackage{algorithm,algorithmic,nicefrac}

\newcommand\modl{\mathfrak{M}}
\newcommand{\by}{\boldsymbol{y}}     
\newcommand\BE{\mathbb{E}}
\newcommand{\prob}{\mathbb{P}}
\newcommand{\esp}{\mathbb{E}}

\begin{document}

\begin{frontmatter}
\title{Likelihood-free Model Choice}
\author{Jean-Michel Marin${}^{1,2}$, Pierre Pudlo${}^{2,3}$, Arnaud Estoup${}^{4}$ and Christian Robert${}^{5,6}$}
\affiliation{
${}^1$IMAG, Universit\'e de Montpellier;
${}^2$IBC, Universit\'e de Montpellier; \\
${}^3$I2M, Aix-Marseille Universit\'e;
${}^4$ CBGP, INRA, Montpellier; \\
${}^5$Universit\'e Paris Dauphine and 
${}^6$University of Warwick}
\begin{abstract}
This document is an invited chapter covering the specificities of ABC model choice, intended for the incoming {\em Handbook of
ABC} by Sisson, Fan, and Beaumont (2017). Beyond exposing the potential pitfalls of ABC approximations to posterior probabilities,
the review emphasizes mostly the solution proposed by \cite{pudlo:etal:2016} on the use
of random forests for aggregating summary statistics and for estimating the posterior probability of the most likely
model via a secondary random forest.
\end{abstract}
\end{frontmatter}

\maketitle

\newcommand\bY{\mathbf{Y}}
\newcommand\yobs{\by^\text{obs}}
\newcommand\sobs{s^\text{obs}}
\newcommand\MF{\mathfrak{M}}
\newcommand\BP{\mathbb{P}}
\newcommand{\knn}{$k$-nn}

\chapter{Likelihood-free model choice}

\section{Introduction}

As it is now hopefully clear from earlier chapters in this book, there exist several ways to set
ABC methods firmly within the Bayesian framework. The method has now gone a very long way from the ``trick" of
the mid 1990's \citep{tavare:balding:griffith:donnelly:1997,pritchard:seielstad:perez:feldman:1999},
where the tolerance acceptance condition
\[
d(\by,\yobs)\le\epsilon
\]
was a crude practical answer to the impossibility to wait for the event $d(\by,\yobs)=0$ associated with exact simulations
from the posterior distribution \citep{rubin:1984}. Not only do we now enjoy theoretical convergence guarantees
\citep{blum:2010,fearnhead:prangle:2012,biau:etal:2015} as the computing power grows to infinity, but we also benefit
from new results that set actual ABC implementations, with their finite computing power and strictly positive
tolerances, within the range of other types of inference \citep{wood:2010,wilkinson:2013,wilkinson:2014}. 
ABC now stands as an inference method that is justifiable on its own ground. This approach may be the only solution
available in complex settings such as those originally tackled in population genetics 
\citep{tavare:balding:griffith:donnelly:1997,pritchard:seielstad:perez:feldman:1999},
unless one engages into more perilous approximations. The conclusion of this evolution towards mainstream Bayesian
inference is quite comforting about the role ABC can play in future computational developments, but this trend is far
from delivering the method a blank confidence check in that some implementations of it will alas fail to achieve consistent inference.

Model choice is actually a fundamental illustration of how much ABC
can err away from providing a proper inference when sufficient care is not properly taken.
This issue is even more relevant when one considers that ABC is
used a lot---at least in population genetics---for the comparison and hence the validation of scenarios that are constructed
based on scientific hypotheses. The more obvious difficulty in ABC model choice is
indeed conceptual rather than computational in that the choice of an inadequate vector of summary statistics
may produce an inconsistent inference \citep{robert:cornuet:marin:pillai:2011} about
the model behind the data. Such an inconsistency cannot be overcome with more powerful computing tools. Existing
solutions avoiding the selection process within a pool of summary statistics are limited to specific problems and
difficult to calibrate.

Past criticisms of ABC from the outside have been most virulent about this aspect, even though not always pertinent
(see, e.g., \cite{templeton:2008,templeton:2010} for an extreme example). It is therefore paramount that the inference
produced by an ABC model choice procedure be validated on the most general possible basis for the method to become
universally accepted. As we discuss in this chapter, reflecting our evolving perspective on the matter,  there are two
issues with the validation of ABC model choice: (a) is it not easy to select a good set of summary statistics
(b) even selecting a collection of summary statistics that lead to a convergent Bayes factor
 may produce a poor approximation at the
practical level. 

As a warning, we note here that this chapter does not provide a comprehensive survey of the literature on ABC model choice,
neither about the foundations \citep[see][]{grelaud:marin:robert:rodolphe:tally:2009,toni:stumpf:2010} and more recent
proposals \citep[see][]{barnes:filippi:stumpf:thorne:2012,prangle:etal:2014,barthelme:chopin:2014},
nor on the wide range of applications of the ABC model choice methodology to specific problems as in, e.g., 
\cite{beaumont:2008,cornuet:ravigne:estoup:2010}.

After introducing standard ABC model choice techniques, we discuss the curse of insufficiency. Then, we present the ABC random
forest strategy for model choice and consider first a toy example and, at the end, a human population genetics example.

\section{Simulate only simulate}

The implementation of ABC model choice should not deviate from the original principle at the core of
ABC, in that it proceeds by treating the unknown model index $\mathfrak{M}$ as an extra parameter
with an associated prior, in accordance with standard Bayesian analysis. An algorithmic
representation associated with the choice of a summary statistic $S(\cdot)$ is thus as
follows:

\begin{footnotesize}
\begin{algorithm}[H]
\caption{standard ABC model choice}
\begin{algorithmic}
\FOR {$i=1$ to $N$}
  \STATE Generate $\modl$ from the prior $\pi(\modl)$
  \STATE Generate $\theta$ from the prior $\pi_{\modl}(\theta)$
  \STATE Generate $\by$ from the model $f_{\modl}(\by|\theta)$
  \STATE Set $\modl^{(i)}=\modl$, $\theta^{(i)}=\theta$ and $s^{(i)}=S(\by)$
\ENDFOR
\RETURN the values $\modl^{(i)}$ associated with the $k$ smallest distances
$d\big(s^{(i)},S(\yobs)\big)$ 
\end{algorithmic}\label{algo:ABCMoo}
\end{algorithm}
\end{footnotesize}

In this presentation of the algorithm, the calibration of the tolerance $\varepsilon$ for ABC model choice is
expressed as a $k$-nearest neighbours (\knn) step, following the validation of ABC in this format
by \cite{biau:etal:2015}, and the observation that the tolerance level is chosen this way in practice.
Indeed, this standard strategy ensure a given number of accepted simulations is produced. While the \knn~method can be used
towards classification and hence model choice, we will take advantage of different machine learning tools in
Section \ref{sec:mach2}.  In general the accuracy of a \knn~method heavily depends on the value of
$k$, which must be calibrated, as illustrated in \cite{pudlo:etal:2016}. Indeed, while the primary
justification of ABC methods is based on the ideal case when $\epsilon\approx 0$, hence $k$ should be taken ``as small
as possible", more advanced theoretical analyses of its non-parametric convergence properties led
to conclude that $\epsilon$ had to be chosen away from zero for a given sample size
\citep{blum:2010,fearnhead:prangle:2012,biau:etal:2015}. Rather than resorting to non-parametric approaches to the
choice of $k$, which are based on asymptotic arguments, \cite{pudlo:etal:2016} rely on an empirical calibration of $k$
using the whole simulated sample known as the reference table to derive the error rate as a function of $k$.

Algorithm \ref{algo:ABCMoo} thus returns a sample of model indices that serves as
an approximate sample from the posterior distribution $\pi(\modl|\yobs)$ and provides an estimated
version via the observed frequencies. In fact, the posterior probabilities can be written as
the following  conditional expectations
\[
\prob\big(\modl = m \big| S(\bY) =s\big) = 
\esp\big(\mathbf 1_{\{\modl = m\}} \big| S(\bY)=s\big).
\]
Computing these conditional expectation based on iid draws from the distribution of
$(\modl, S(\bY))$ can be interpreted as a regression problem in which the response is the indicator of
whether or not the simulation comes from model $m$ and the covariates are the summary statistics. The iid draws
constitute the reference table, which also is the training database for machine learning methods. The process used in the
above ABC Algorithm \ref{algo:ABCMoo} is a \knn method if one approximates the posterior by the frequency of $m$ 
among the $k$ nearest simulations to $s$. The proposals of \cite{grelaud:marin:robert:rodolphe:tally:2009}
and \cite{toni:etal:2009} for ABC model choice are exactly in that vein.

Other methods can be implemented to better estimate $\prob\big(\modl = m \big| S(\bY) =s\big)$ from
the reference table, the training database of the regression method. For instance, Nadaraya-Watson
estimators are weighted averages of the responses, where weights are non-negative decreasing functions (or kernels)
of the distance $d(s^{(i)},s)$. The regression method commonly used (instead of \knn) is a local 
regression method, with a multinomial link, as proposed by \cite{fagundes:etal:2007} or by
\cite{cornuet:ravigne:estoup:2010}: local regression procedures fit a linear model on simulated pairs
$(\modl^{(i)},s^{(i)})$ in a neighbourhood of $s$. The multinomial link ensures that the vector of
probabilities has entries between $0$ and $1$ and sums to $1$. However, local regression can prove
computationally expensive, if not intractable, when the dimension of the covariate increases. Therefore,
\cite{estoup:etal:2012} proposed a dimension reduction technique based on linear discriminant analysis (an exploratory
data analysis technique that projects the observation cloud along axes that maximise the discrepancies between groups, see
\cite{hastie:tibshirani:friedman:2009}), which produces to a summary statistic of dimension $M-1$.

 \begin{algorithm}[H]
   \caption{\sffamily local logistic regression ABC model choice}
   \label{algo:reglo}
   \begin{algorithmic}\sffamily
\STATE Generate $N$ samples $\big(\modl^{(i)},s^{(i)}\big)$ as in Algorithm \ref{algo:ABCMoo}
\STATE Compute weights $\omega_i=K_\mathfrak{h}(s^{(i)}-S(y^\text{obs}))$ where $K$ is a kernel density and $\mathfrak{h}$ is its
bandwidth estimated from the sample $\big(s^{(i)}\big)$
\STATE Estimate the probabilities $\prob\big(\modl = m \big| s\big)$ by a logistic link based on the covariate $s$
from the weighted data $\big(\modl^{(i)},s^{(i)},\omega_i\big)$
   \end{algorithmic}
\end{algorithm}

Unfortunately, all regression procedures given so far suffer from a curse of dimensionality: they are 
sensitive to the number of covariates, i.e., the dimension of the vector of summary
statistics. Moreover, as detailed in the following sections, any improvements in the regression
method do not change the fact that all these methods aim at approximating $\prob\big(\modl = m \big|
S(\bY) =s\big)$ as a function of $s$ and use this function at $s=s^\text{obs}$, while caution and
cross-checking might be necessary to validate $\prob\big(\modl = m \big|
S(\bY) =s^\text{obs}\big)$ as an approximation of $\prob\big(\modl = m \big| \bY =\yobs\big)$.

A related approach worth mentioning here is the Expectation Propagation ABC (EP-ABC)
algorithm of \cite{barthelme:chopin:2014}, which also produces an
approximation of the evidence associated with each model under comparison. Without getting into details, the
expectation-propagation approach of \cite{minka:lafferty:2002,seeger:2005} approximates the posterior distribution by a
member of an exponential family, using an iterative and fast moment-matching process that takes only a component of the likelihood
product at a time. When the likelihood function is unavailable, \cite{barthelme:chopin:2014} propose to instead rely on empirical
moments based on simulations of those fractions of the data. The algorithm includes as a side product an estimate of the
evidence associated with the model and the data, hence can be exploited for model selection and posterior probability
approximation. On the positive side, the EP-ABC is much faster than a standard ABC scheme, does not always resort to summary
statistics, or at least to global statistics, and is appropriate for ``big data" settings where the whole data cannot be
explored at once. On the negative side, this approach has the same degree of validation as variational Bayes methods
\cite{jaakkola:jordan:2000}, which means converging to a proxy posterior that is at best optimally close to the genuine
posterior within a certain class, requires a meaningful decomposition of the likelihood into blocks which can be simulated, calls for
the determination of several tolerance levels, is critically dependent on calibration choices, has
no self-control safety mechanism and requires identifiability of the models' underlying parameters. 
Hence, while EP-ABC can be considered for conducting model selection, there is no
theoretical guarantee that it provides a converging approximation of the evidence, while the implementation on realistic
models in population genetics seems out of reach.

\section{The curse of insufficiency}

The paper \citep{robert:cornuet:marin:pillai:2011} issued a warning that ABC approximations to
posterior probabilities cannot always be trusted in the double sense that (a) they stand away from the
genuine posterior probabilities (imprecision) and (b) they may even fail to converge to a Dirac distribution on the
true model as the size of the observed dataset grows to infinity (inconsistency). Approximating
posterior probabilities via an ABC algorithm means using the frequencies of acceptances of
simulations from each of those models. We assumed in Algorithm 1 the use of a common summary
statistic (vector) to define the distance to the observations as otherwise the comparison between
models would not make sense. 
This point may sound anticlimactic since the same feature occurs for point estimation, where the ABC
estimator is an estimate of $\BE[\theta|S(\yobs)]$. Indeed, all ABC approximations
rely on the posterior distributions knowing those summary statistics, rather than knowing the whole
dataset. When conducting point estimation with insufficient statistics, the information content is
necessarily degraded.  The posterior distribution is then different from the true posterior but, at
least, gathering more observations brings more information about the parameter (and convergence when
the number of observations goes to infinity), unless one uses only ancillary statistics. However,
while this information impoverishment only has consequences in terms of the precision of the
inference for most inferential purposes, it induces a dramatic arbitrariness in the construction of
the Bayes factor. To illustrate this arbitrariness, consider the case of starting from a 
statistic $S(x)$ sufficient for both models. Then, by the factorisation theorem, the true likelihoods factorise
as
\[
f_1(x|\theta) = g_1(x) \pi_1(S(x)|\theta) \quad\text{and}\quad 
f_2(x|\theta) = g_2(x) \pi_2(S(x)|\theta)
\]
resulting in a true Bayes factor equal to
\begin{equation}\label{eq:wronBF}
B_{12}(x) = \dfrac{g_1(x)}{g_2(x)}\,B^S_{12}(x)
\end{equation}
where the last term, indexed by the summary statistic $S$, is the limiting (or Monte Carlo error-free) version of the ABC Bayes
factor. In the more usual case where the user cannot resort to a sufficient statistic, the ABC Bayes factor may diverge
one way or another as the number of observations increases. A notable exception is the case of Gibbs random fields where
\cite{grelaud:marin:robert:rodolphe:tally:2009} have shown how to derive inter-model sufficient statistics, beyond the
raw sample.  This is related to the less pessimistic paper of \cite{didelot:everitt:johansen:lawson:2011}, also
concerned with the limiting behaviour for the ratio \eqref{eq:wronBF}.  Indeed, these authors reach the opposite
conclusion from ours, namely that the problem can be solved by a sufficiency argument. Their point is that, when
comparing models within exponential families (which is the natural realm for sufficient statistics), it is always
possible to build an encompassing model with a sufficient statistic that remains sufficient across models.

However, apart from examples where a tractable sufficient summary statistic is identified, one cannot easily compute a
sufficient summary statistic for model choice and this results in a loss of information, when compared with the exact
inferential approach, hence a wider discrepancy between the exact Bayes factor and the quantity produced by an ABC
approximation. When realising this conceptual difficulty, the authors of \cite{robert:cornuet:marin:pillai:2011} felt it
was their duty to warn the community about the dangers of this approximation, especially when considering the rapidly
increasing number of applications using ABC for conducting model choice or hypothesis testing. Another argument in
favour of this warning is that it is often difficult in practice to design a summary statistic that is informative about
the model.

Let us signal here that a summary selection approach purposely geared towards model selection can be found in
\cite{barnes:filippi:stumpf:thorne:2012}. Let us stress in and for this section that the said method similarly suffers from the above
curse of dimensionality. Indeed, the approach therein is based on an estimate of Fisher's information contained in the
summary statistics about the pair $(\modl,\theta)$ and the correlated search for a subset of those summary statistics
that is (nearly) sufficient. As explained in the paper, this approach implies that the resulting summary statistics are
also sufficient for parameter estimation within each model, which obviously induces a dimension inflation in the
dimension of the resulting statistic, in opposition to approaches focussing solely on the selection of summary
statistics for model choice, like \cite{prangle:etal:2014} and \cite{cornuet:etal:2014}.

We must also stress that, from a model choice perspective, the vector made of the (exact!) posterior probabilities of
the different models obviously constitutes a Bayesian sufficient statistics of dimension $M-1$, but this vector is
intractable precisely in cases where the user has to resort to ABC approximations. Nevertheless, this remark is exploited
in \cite{prangle:etal:2014} in a two-stage ABC algorithm. The second stage of the algorithm is ABC model choice with 
summary statistics equal to approximation of the posterior probabilities. Those approximations are computed as ABC
solutions at the first stage of the algorithm. Despite the conceptual attractiveness of this approach, which relies on a
genuine sufficiency result, the approximation of the posterior probabilities given by the first stage of the algorithm 
directly rely on the choice of a particular set of summary statistics, which brings us back to the original issue of
trusting an ABC approximation of a posterior probability. 

There therefore is a strict loss of information in using ABC model choice, due to the call both to insufficient
statistics and to non-zero tolerances (or a imperfect recovery of the posterior probabilities with a regression
procedure).

\begin{figure}
\centering
\includegraphics[width=.59\textwidth]{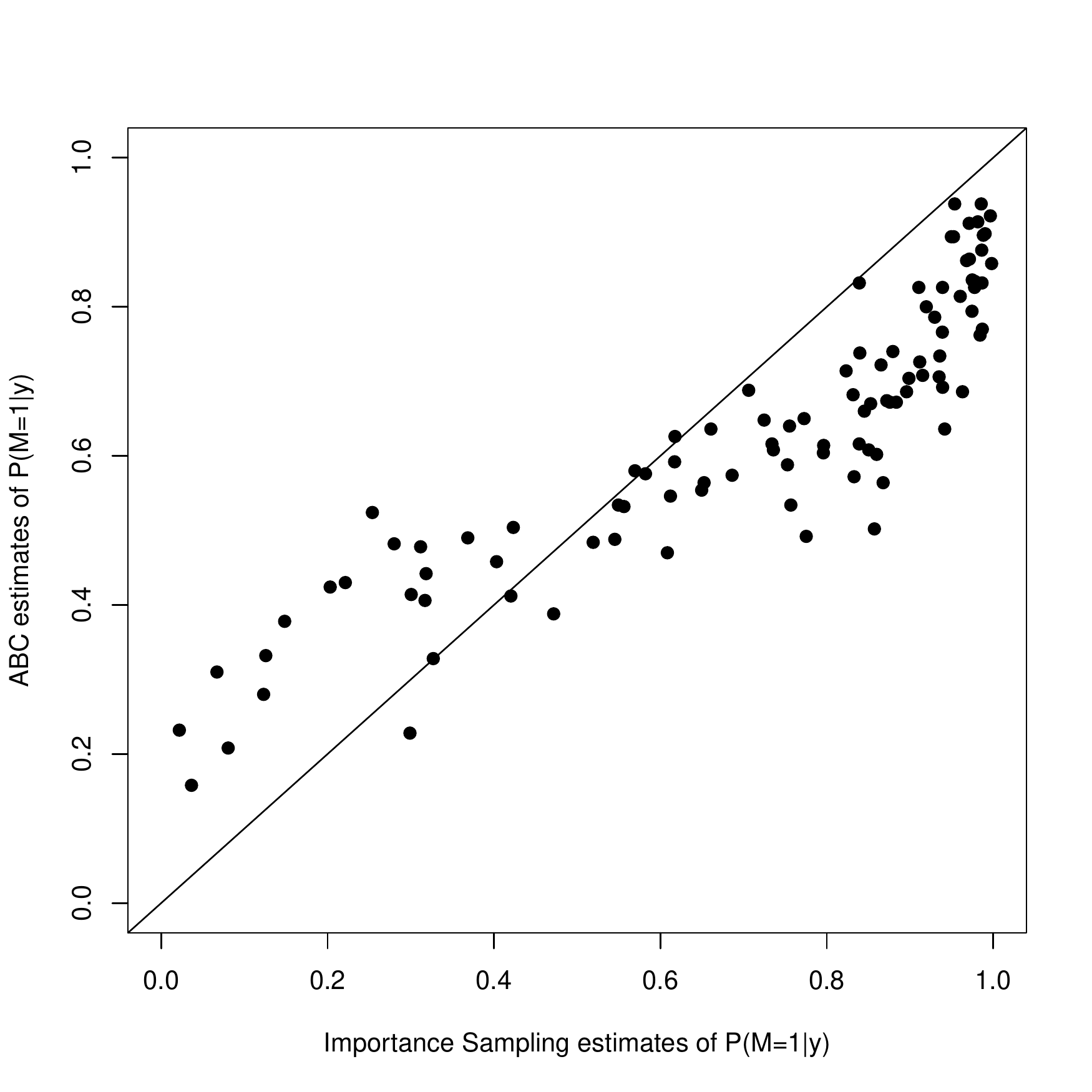}
\caption{\label{fig:res11}Comparison of importance sampling {\em (first axis)} and ABC {\em (second axis)} 
estimates of the posterior probability of scenario 1 in the first population genetic experiment, using 24 
summary statistics. {\em (Source: \cite{robert:cornuet:marin:pillai:2011})}}
\end{figure}

\subsection{Some counter-examples}

Besides a toy example opposing Poisson and Geometric distributions to point out the potential
irrelevance of the Bayes factor based on poor statistics, \cite{robert:cornuet:marin:pillai:2011}
goes over a realistic population genetic illustration, where two evolution scenarios involving three
populations are compared, two of those populations having diverged 100 generations ago and the third one
resulting from a recent admixture between the first two populations (scenario 1) or simply diverging
from population 1 (scenario 2) at the same date of 5 generations in the past. In scenario 1, the
admixture rate is 0.7 from population 1.  Simulated datasets (100) of the same size as in experiment
1 (15 diploid individuals per population, 5 independent micro-satellite loci) were generated
assuming an effective population size of 1000 and a mutation rate of 0.0005. In this experiment,
there are six parameters (provided with the corresponding priors): the admixture rate
($\mathcal{U}[0.1,0.9]$), three effective population sizes ($\mathcal{U}[200,2000]$), the time of
admixture/second divergence ($\mathcal{U}[1,10]$) and the date of the first divergence
($\mathcal{U}[50,500]$). While costly in computing time, the posterior probability of a scenario can
be estimated by importance sampling, based on 1000 parameter values and 1000 trees per parameter
value, thanks to the modules of \cite{stephens:donnelly:2000}. The ABC approximation is produced by
DIYABC \citep{cornuet:etal:2008}, based on a reference sample of two million parameters and 24
summary statistics. The result of this experiment is shown on Figure \ref{fig:res11}, with a clear
divergence in the numerical values despite stability in both approximations. Taking the importance
sampling approximation as the reference value, the error rates in using the ABC approximation to
choose between scenarios 1 and 2 are 14.5\%~and 12.5\%~(under scenarios 1 and 2),
respectively. Although a simpler experiment with a single parameter and the same 24 summary
statistics shows a reasonable agreement between both approximations, this result comes as an additional
support to our warning about a blind use of ABC for model selection.  The corresponding simulation
experiment was quite intense, as, with 50 markers and 100 individuals, the product likelihood
suffers from an enormous variability that 100,000 particles and 100 trees per locus have trouble
addressing despite a huge computing cost.

\begin{figure}
 \centerline{\includegraphics[width=9truecm]{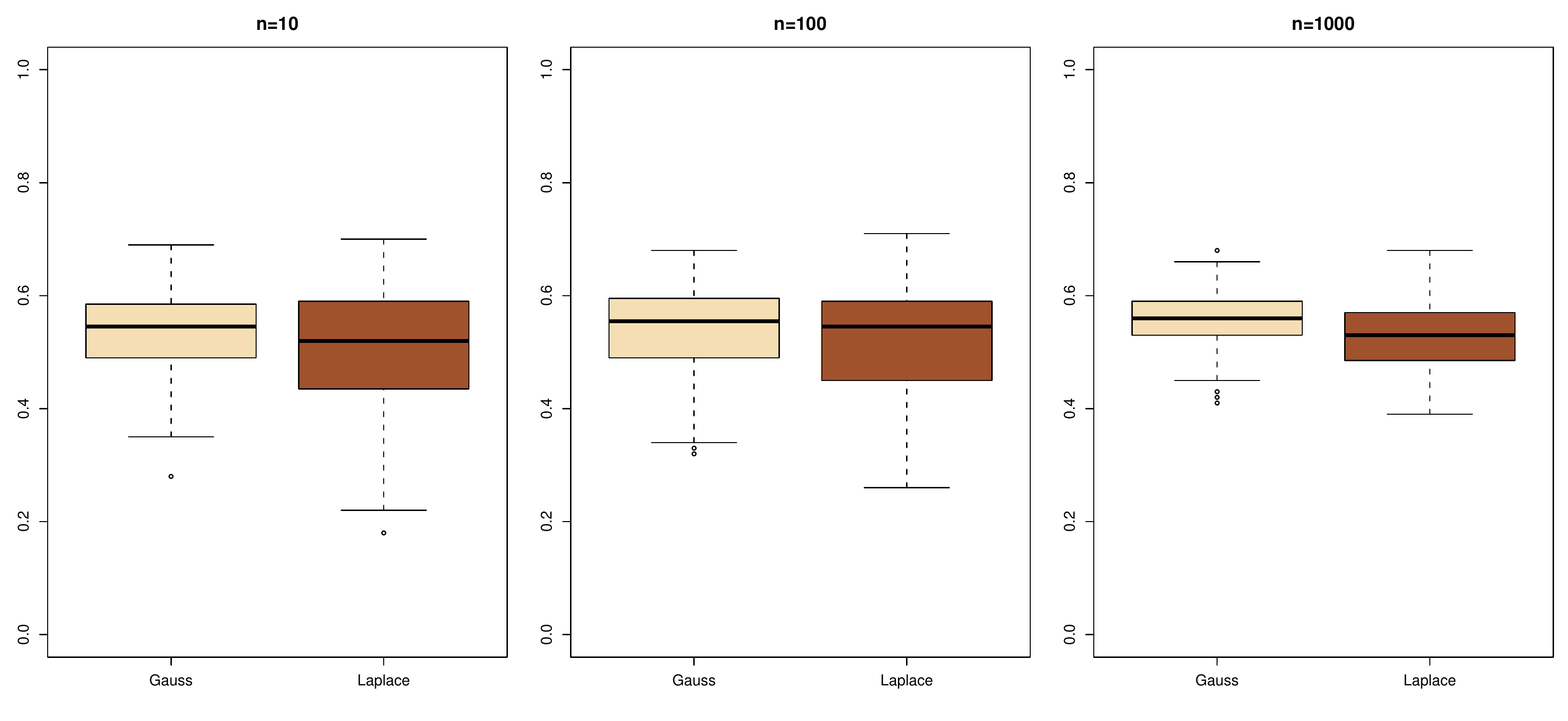}}
\caption{Comparison of the range of the ABC posterior probability that data is from a normal model with unknown mean $\theta$
when the data is made of $n=10,100,1000$ observations
{\em (left, centre, right, resp.)} either from a Gaussian {\em (lighter)} or
Laplace distribution {\em (darker)} and when the ABC summary statistic is
made of the empirical mean, median, and variance. The ABC algorithm generates
$10^4$ simulations ($5,000$ for each model) from the prior $\theta\sim \mathcal{N}(0,4)$
and selects the tolerance $\epsilon$ as the $1\%$ distance quantile over those simulations.
{\em (Source: \cite{marin:pillai:robert:rousseau:2011}.)}}
\label{fig:norlap}
\end{figure}

An example is provided in the introduction of the paper \cite{marin:pillai:robert:rousseau:2011}, sequel to
\cite{robert:cornuet:marin:pillai:2011}. The setting is one of a
comparison between a normal $\by\sim\mathcal{N}(\theta_1,1)$ model and a double exponential
$\by\sim\mathcal{L}(\theta_2,1/\sqrt{2})$ model\footnote{The double exponential distribution is also
  called the Laplace distribution, hence the notation $\mathcal{L}(\theta_2,1/\sqrt{2})$, with mean
  $\theta_2$ and variance one.}. The summary statistics used in the corresponding ABC algorithm are
the sample mean, the sample median and the sample variance. Figure \ref{fig:norlap} exhibits the
absence of discrimination between both models, since the posterior probability of the normal
model converges to a central value around $0.5$-$0.6$ when the sample size grows, irrelevant of the
true model behind the simulated datasets.

\subsection{Still some theoretical guarantees}

Our answer to the (well-received) above warning is provided in \cite{marin:pillai:robert:rousseau:2011}, which deals
with the evaluation of summary statistics for Bayesian model choice.  The main result states that, under some Bayesian
asymptotic assumptions, ABC model selection only depends on the behaviour of the mean of the summary statistic under
both models. The paper establishes a theoretical framework that leads to demonstrate consistency of the ABC Bayes factor
under the constraint that the ranges of the expected value of the summary statistic under both models do not intersect.
An negative result is also given in \cite{marin:pillai:robert:rousseau:2011}, which mainly states that, whatever the
observed dataset, the ABC Bayes factor selects the model having the smallest effective dimension when the assumptions
do not hold.

\begin{figure}
 \centerline{\includegraphics[width=9truecm]{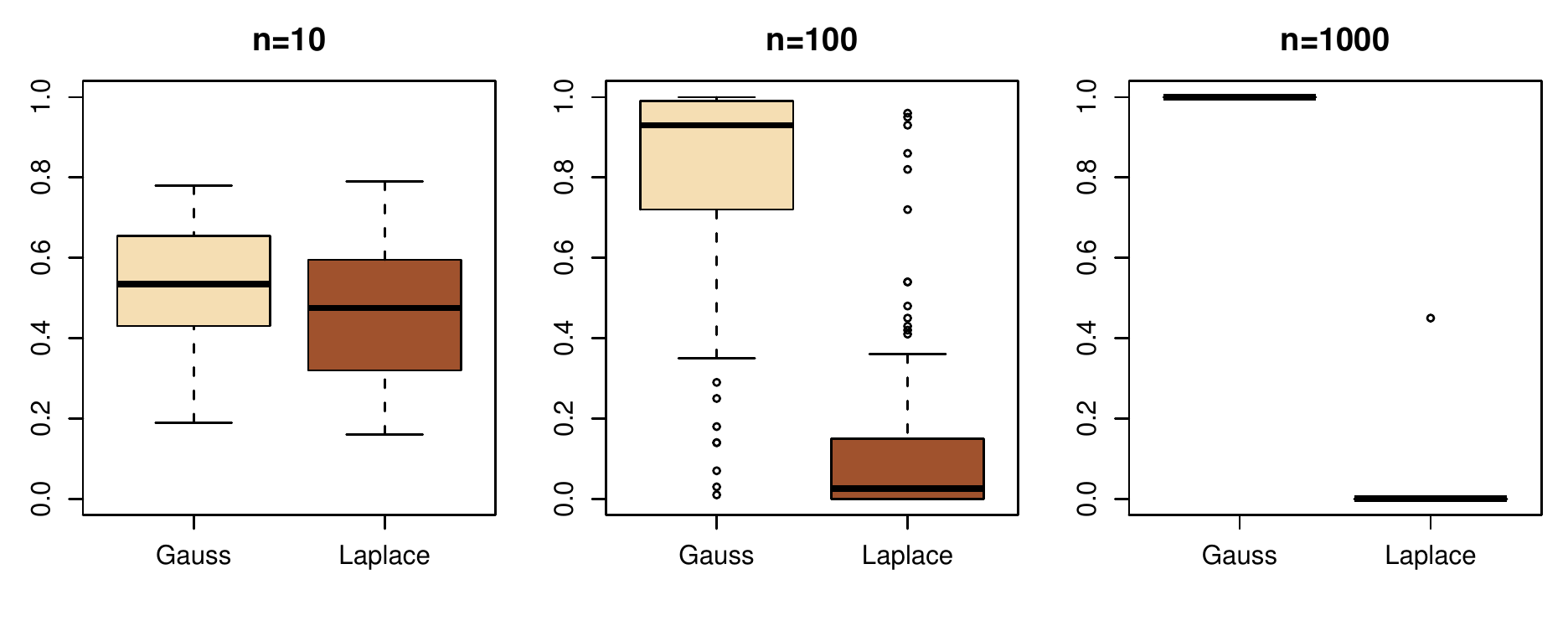}}
\caption{Same representation as Figure \ref{fig:norlap} when using
the median absolute deviation of the sample as its sole summary statistic.
{\em (Source: \cite{marin:pillai:robert:rousseau:2011}.)}}
\label{fig:norlip}
\end{figure}

The simulations associated with the paper were straightforward in that (a) the setup
compares normal and Laplace distributions with different summary statistics (inc.~the median
absolute deviation), (b) the theoretical results told what to look for, and (c) they did very
clearly exhibit the consistency and inconsistency of the Bayes factor/posterior probability
predicted by the theory. Both boxplots shown here on Figures \ref{fig:norlap} and \ref{fig:norlip}
show this agreement: when using (empirical) mean, median, and variance to compare normal and Laplace
models, the posterior probabilities do not select the true€ model but instead aggregate near a
fixed value. When using instead the median absolute deviation as summary statistic, the posterior
probabilities concentrate near one or zero depending on whether or not the normal model is the true
model.

It may be objected to such necessary and sufficient conditions that Bayes factors simply are inappropriate for
conducting model choice, thus making the whole derivation irrelevant. This foundational perspective is an arguable
viewpoint \citep{evans:2015}.  However, it can be countered within the Bayesian paradygm by the fact that Bayes factors
and posterior probabilities are consistent quantities that are used in conjunction with ABC in dozens of genetic papers.
Further arguments are provided in the various replies to both of Templeton's radical criticisms
\citep{templeton:2008,templeton:2010}. That more empirical and model-based assessments also are available is quite
correct, as demonstrated in the multicriterion approach of \cite{ratmann:andrieu:wiujf:richardson:2009}. This is simply
another approach, not followed by most geneticists so far.

A concluding remark about \cite{marin:pillai:robert:rousseau:2011} is that, while the main bulk of
the paper is theoretical, it does bring an answer that the mean ranges of the summary statistic
under each model must not intersect if they are to be used for ABC model choice. In addition, while the theoretical
assumptions therein are not of the utmost relevance for statistical practice, the paper includes recommendations on how to conduct
a $\chi^2$ test on the difference of the means of a given summary statistics under both models, towards assessing
whether or not this summary is acceptable.

\section{Selecting the MAP model via machine learning}\label{sec:mach2}

The above sections provide enough arguments to feel less than confident in the outcome of a standard ABC model choice
algorithm \ref{algo:ABCMoo}, at least in the numerical approximation of the probabilities 
$\BP(\MF=m|S(\bY)=s^\text{obs})$ and in their connection with the genuine posterior probabilities
$\BP(\MF=m|\bY=\yobs)$. There are indeed three levels of approximation errors in such quantities, one due to the Monte
Carlo variability, one due to the non-zero ABC tolerance or, more generally to the error committed by the regression
procedure when estimating the conditional expected value, and one due to the curse of insufficiency. While the
derivation of a satisfying approximation of the genuine $\BP(\MF=m|\bY=\yobs)$ seems beyond our reach, we present
below a novel approach to both construct the most likely model and approximate
$\BP(\MF=m|S(\bY)=s^\text{obs})$ for the most likely model, based on the machine learning tool of random forests.

\subsection{Reconsidering the posterior probability estimation}

Somewhat paradoxically, since the ABC approximation to posterior probabilities of a collection of models is delicate,
\cite{pudlo:etal:2016} support inverting the order of selection of the a posteriori most probable model and of
approximation of its posterior probability, using the alternative tool of random forests for both goals. The reason for
this shift in order is that the rate of convergence of local regression procedure such as \knn\ or the local regression with multinomial
link heavily depends on the dimension of the covariates (here the dimension of the summary statistic). 
Thus, since the primary goal of ABC model choice is to select the most appropriate model, both \cite{stoehr:pudlo:cucala:2015} and \cite{pudlo:etal:2016}
argue that one does not need to correctly approximating the probability $$\BP(\MF=m|S(\bY)\approx s^\text{obs})$$ when
looking for the most probable model in the sense of $$\BP(\MF=m|\bY=\yobs)$$ probability.  \cite{stoehr:pudlo:cucala:2015}
stresses that selecting the most adequate model for the data at hand as the maximum a posteriori (MAP) model index 
is a classification issue, which proves to be a significantly easier inference problem than estimating a regression function
\citep{hastie:tibshirani:friedman:2009,devroye:gyorfi:lugosi:1996}. This is the reason why \cite{stoehr:pudlo:cucala:2015} adapt
the above Algorithm~\ref{algo:ABCMoo} by resorting to a \knn\ classification procedure, which sums up as
returning the most frequent (or majority rule) model index among the $k$ simulations nearest to the observed dataset,
nearest in the subspace of the summary statistics. Indeed, generic classification aims at forecasting a
variable $\MF$ taking a finite number of values, $\{1,\ldots, M\}$, based on a vector of covariates
$S=(S_1,\ldots, S_d)$. The Bayesian approach to classification stands in using a training
database $(m^i, s^i)$ made of independent replicates of the pair $(\MF,S(\bY))$ that are simulated from the prior predictive distribution.
The connection with ABC model choice is that the later predicts a model index, $\MF$, from
the summary statistic $S(\bY)$. Simulations in the ABC reference table can thus be envisioned as creating a learning
database from the prior predictive that trains the classifier.

\cite{pudlo:etal:2016} widen the paradigm shift undertaken in \cite{stoehr:pudlo:cucala:2015}, as they use a machine
learning approach to the selection of the most adequate model for the data at hand and exploit this tool to derive an
approximation of the posterior probability of the selected model. The classification procedure chosen by
\cite{pudlo:etal:2016} is the technique of Random Forests (RFs)  \citep{breiman:2001}, which constitutes a trustworthy and
seasoned machine learning tool, well adapted to complex settings as those found in ABC settings. The approach further
requires no primary selection of a small subset of summary statistics, which allows for an automatic input of summaries
from various sources, including softwares like DIYABC \citep{cornuet:etal:2014}.
At a first stage, a RF is constructed from the reference table to predict the model index and applied to the data at hand to return a MAP estimate.
At a second stage, an additional RF is constructed for explaining the selection error of the MAP estimate, based
on the same reference table. When applied to the observed data, this secondary random forest produces an estimate of the
posterior probability of the model selected by the primary RF, as detailed below, following \cite{pudlo:etal:2016}.

 \subsection{Random forests construction}

A RF aggregates a large number of classification trees by adding for each tree a randomisation step to the Classification And
Regression Trees (CART) algorithm \citep{breiman:friedman:stone:olshen:1984}.
Let us recall that this algorithm produces a binary classification tree that partitions the covariate space towards a prediction
of the model index. In this tree, each binary node is partitioning
the observations via a rule of the form $S_j<t_j$, where $S_j$ is one of the summary statistics and $t_j$ is chosen
towards the minimisation of an heterogeneity index. For instance, \cite{pudlo:etal:2016} uses the Gini criterion
\citep{hastie:tibshirani:friedman:2009}. A CART tree is built based on a learning table and it is then applied to the
observed summary statistic $\sobs$, predicting the model index by following a path that applies these binary rules
starting from the tree root and returning the label of the tip at the end of the path.

The randomisation part in RF produces a large number of distinct CART trees by (a) using for each tree a
bootstrapped version of the learning table on a bootstrap sub-sample of size
$N_\text{boot}$ and (b) selecting the summary statistics at each node from a random subset of the available summaries.
The calibration of a RF thus involves three quantities:  
\begin{itemize}
\item[--] $B$, the number of trees in the forest, 
\item[--] $n_\text{try}$, the number of covariates randomly sampled at each node by the randomised CART, and 
\item[--] $N_\text{boot}$, the size of the bootstraped sub-sample. 
\end{itemize}
The so-called out-of-bag error associated with an RF is the average number of times a point from the learning
table is wrongly allocated, when averaged over trees that exclude this point from the bootstrap sample.

The way \cite{pudlo:etal:2016} builds a random forest classifier given a collection of statistical models is
to start from an ABC reference table including a set of simulation records made of model indices,
parameter values and summary statistics for the associated simulated data. This table then serves as
training database for a random forest that forecasts model index based on the summary statistics.

\begin{algorithm}[H]
   \caption{\sffamily random forest ABC model choice}
   \begin{algorithmic}\sffamily
\STATE Generate $N$ samples $\big(\modl^{(i)},s^{(i)}\big)$ as in Algorithm \ref{algo:ABCMoo} (the reference table) 
\STATE Construct $N_\text{tree}$ randomized CART which predict the model indices using the summary statistics
    \FOR{$b=1$ \TO $N_\text{tree}$}
    \STATE \textbf{draw} a bootstrap sub-sample of size $N_\text{boot}$ from the reference table
    \STATE \textbf{grow} a randomized CART $T_b$
    \ENDFOR
\STATE Determine the predicted indices for $\sobs$ and the trees $\{T_b;b=1,\ldots,N_\text{tree}\}$
\STATE Assign $\sobs$ to an indice (a model) according to a majority vote among the predicted indices 
\end{algorithmic}
\end{algorithm}

\subsection{Approximating the posterior probability of the MAP}

The posterior probability of a model is the natural Bayesian uncertainty quantification \citep{robert:2001} since it is the complement of
the posterior loss associated with a 0--1 loss  $\mathbf 1_{\modl \neq \hat\modl(\sobs)}$
where $\hat\modl(\sobs)$ is the model selection procedure, e.g., the RF outcome described in the above
section. However, for reasons described above, we are unwilling to trust the standard ABC approximation to the posterior 
probability as reported in Algorithm~\ref{algo:ABCMoo}. An initial proposal in \cite{stoehr:pudlo:cucala:2015} is to instead rely on the conditional error rate
induced by the \knn\ classifier knowing $S(\bY)=s^\text{obs}$, namely
\[
\prob\big(\modl \neq \widehat{\modl}(\sobs)  \big| \sobs\big) \,,
\]
where $\widehat{\modl}$ denotes the \knn\ classifier trained on ABC simulations. The above
conditional expected value of $\mathbf 1_{\{ \modl \neq \widehat{\modl}(\sobs) \}}$ is approximated in
\cite{stoehr:pudlo:cucala:2015} with a Nadaraya-Watson estimator on a new set of simulations where
the authors compare the model index $m^{(i)}$ which calibrates the simulation of the pseudo-data $y^{(i)}$,
and the model index $\widehat{\modl}(s^{(i)})$ predicted by the \knn\ approach trained on a first database
of simulations.  However, this first proposal has the major drawback of relying on nonparametric
regression, which deteriorates when the dimension of the summary statistic increases. This local error
also allows for the selection of summary statistics adapted to $\sobs$ but the procedure of
\cite{stoehr:pudlo:cucala:2015} remains constrained by the dimension of the summary statistic,
which typically have to be less than 10.

Furthermore, relying on a large dimensional summary statistic---to bypass, at least partially, the curse of
insufficiency---was the main reason for adopting a classifier such as RFs in \cite{pudlo:etal:2016}. Hence
the authors proposed to estimate the posterior expectation of $\mathbf 1_{\modl \neq \hat\modl(\sobs)}$ as a function of the summary statistics, via
another RF construction.
\begin{align*}
\mathbb{E}[\mathbf 1_{\modl \neq \hat\modl(\sobs)}|\sobs] & = \mathbb{P}[\modl \neq \hat\modl(\sobs)|\sobs]\\
                                                          & = 1-\mathbb{P}[\modl = \hat\modl(\sobs)|\sobs]\,.
\end{align*}
The estimation of $\mathbb{E}[\mathbf 1_{\modl \neq \hat\modl(s)}|s]$ proceeds as follows: 
\begin{itemize}
\item[--] compute the values of $\mathbf 1_{\modl \neq \hat\modl(s)}$ for the trained random forest and all terms in the
reference table;
\item[--] train a second RF regressing $\mathbf 1_{\modl \neq \hat\modl(s)}$ on the same set of summary
statistics and the same reference table, producing a function $\varrho(s)$ that returns a machine learning estimate of
$\mathbb{P}[\modl\ne \hat\modl(s)|s]$;
\item[--] apply this function to the actual observations to produce $1-\varrho(\sobs)$ as an estimate of $\mathbb{P}[\modl
= \hat\modl(\sobs)|\sobs]$.
\end{itemize}

\section{A first toy example}

We consider in this section a simple unidimensional setting with three models
where the marginal likelihoods can be computed in closed form.

Under Model 1, our dataset is a $n$-sample from an Exponential distribution with parameter
$\theta$ (with expectation $1/\theta$) and the corresponding prior distribution on $\theta$ is an Exponential 
distribution with parameter 1.  In this model, given the sample $y=(y_1,\ldots,y_n)$ with $y_i>0$,
the marginal likelihood is given by
$$
m_1(\by)=\Gamma(n+1)\left(1+\sum_{i=1}^n y_i\right)^{-n-1}
$$

Under Model 2, our dataset is a $n$-sample from a Log-Normal distribution with location parameter
$\theta$ and dispersion parameter equal to 1 (which implies an expectation equal to $\exp(\theta+0.5)$).
The prior distribution on $\theta$ is a standard Gaussian distribution. For this model, given the sample
$y=(y_1,\ldots,y_n)$ with $y_i>0$, the marginal likelihood is given by
$$
m_2(\by)=\exp\left[-\left(\sum_{i=1}^n\log(y_i)\right)^2/(2n(n+1))-
\left(\sum_{i=1}^n\log^2(y_i)\right)^2/2\right.
$$
$$\left.+\left(\sum_{i=1}^n\log(y_i)\right)^2/(2n)
-\sum_{i=1}^n\log(y_i)\right]\times (2\pi)^{-n/2}\times (n+1)^{-1/2}
$$

Under Model 3, our dataset is a $n$-sample from a Gamma distribution with parameter $(2,\theta)$
(with expectation $2/\theta$) and the prior distribution on $\theta$ is an Exponential distribution with parameter
1. For this model, given the sample $y=(y_1,\ldots,y_n)$with $y_i>0$, the marginal likelihood is given by
$$
m_3(\by)=\exp\left[\sum_{i=1}^n \log(y_i)\right]\frac{\Gamma(2n+1)}{\Gamma(2)^n}\left(1+\sum_{i=1}^n y_i\right)^{-2n-1}
$$

We consider three summary statistics
$$
\left(\sum_{i=1}^n y_i,\sum_{i=1}^n \log(y_i),
\sum_{i=1}^n \log^2(y_i)\right)\,.
$$
These summary statistics are sufficient not only within each model but also for the model 
choice problem \citep{didelot:everitt:johansen:lawson:2011} and the purpose of this example is not to evaluate
the impact of a loss of sufficiency.

When running ABC, we set $n=20$ for the sample size and generated a reference table containing $29,000$ simulations
(9676 simulations from model 1, 9650 from model 2 and 9674 from model 3).  We further generated an independent test
dataset of size 1,000.  Then, to calibrate the optimal number of neighbours in the standard ABC procedure
\citep{grelaud:marin:robert:rodolphe:tally:2009,toni:etal:2009} we exploited
$1,000$ independent simulations.

\begin{figure}
\centering
\includegraphics[width=0.8\textwidth]{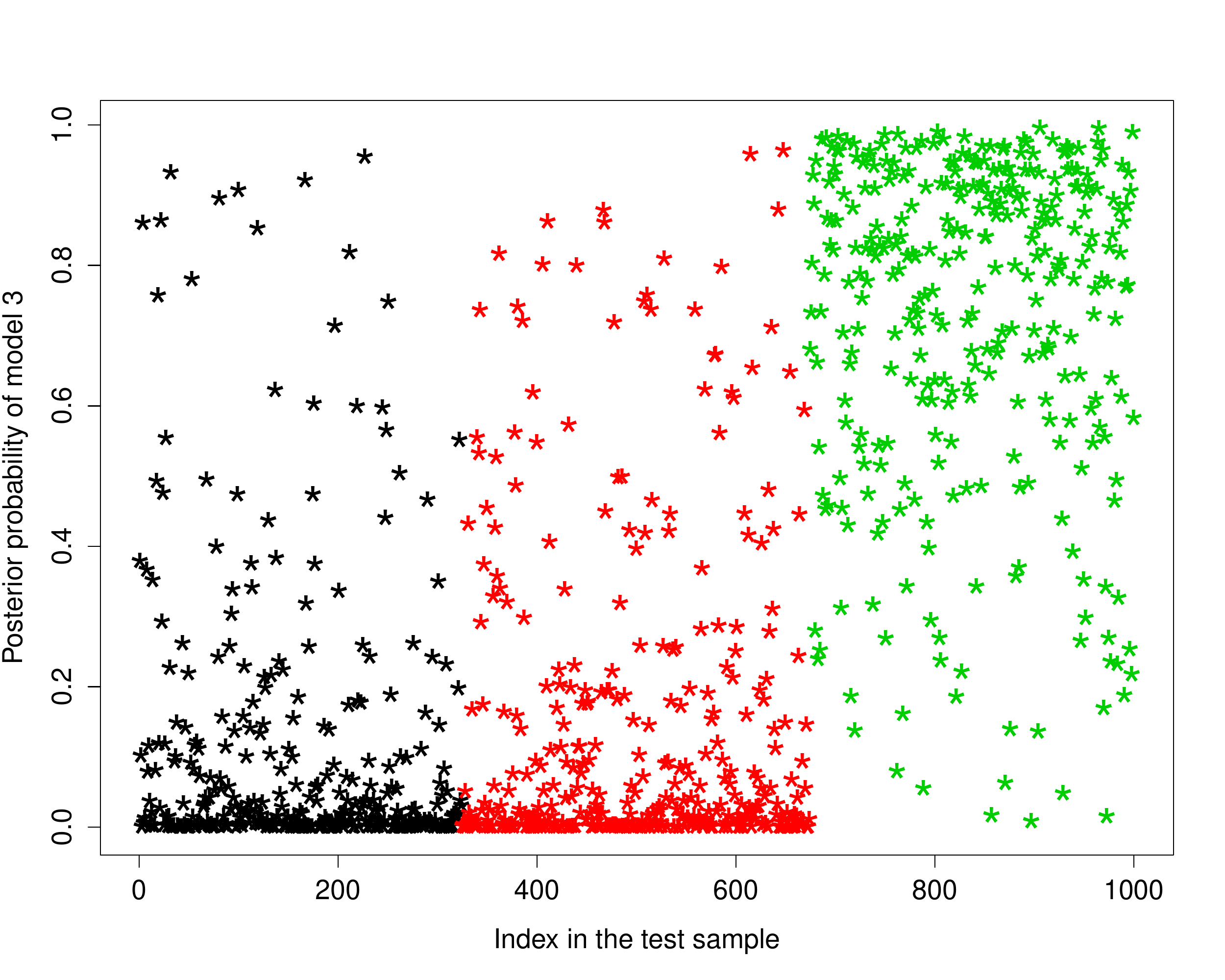}
\caption{\label{fig:toy_post} True posterior probability of Model 3 for each term from the test sample.
Colour corresponds to the true model index: black for Model 1, red for Model 2 and green for Model 3. The terms in the test
sample have been ordered by model index to improve the representation.}
\end{figure}

\begin{figure}
\centering
\includegraphics[width=0.8\textwidth]{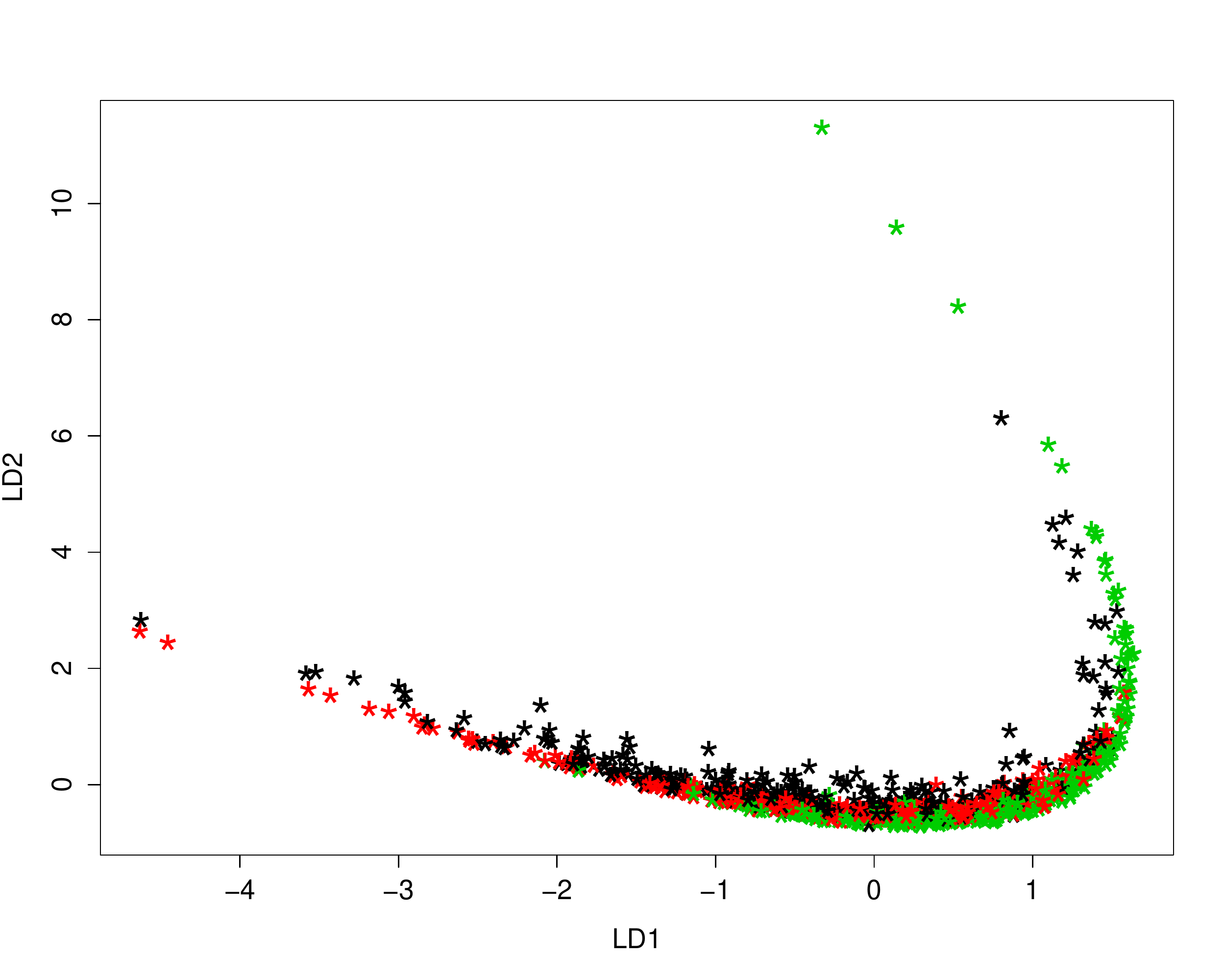}
\caption{\label{fig:toy_LDA} LDA projection along the first two axes of the test dataset, with the same colour code as
in Figure \ref{fig:toy_post}.}
\end{figure}

For each element of the test dataset, as obvious from the above $m_i(\by)$'s we can evaluate
the exact model posterior probabilities. Figure \ref{fig:toy_post} represents the posterior probability of Model 3 for
every simulation, ranked by model index. In addition, Figure \ref{fig:toy_LDA} gives a plot of
the first two LDA projections of the test dataset. Both figures explain why the model choice problem is not easy in this setting.
Indeed, based on the exact posterior probabilities, selecting the model associated with the highest posterior
probability achieves the smallest prior error rate.  Based on the test dataset, we estimate this lower bound as being
around 0.245, i.e., close to 25 \%.

Based on a calibration set of 1,000 simulations, and the above reference table of size 29,000,
the optimal number of neighbours that should be used by the standard ABC model choice procedure,
i.e., the one that minimises the prior error rate, is equal to 20. In this case,
the resulting prior error rate for the test dataset is equal to 0.277.

\begin{figure}
\centering
\includegraphics[width=0.8\textwidth]{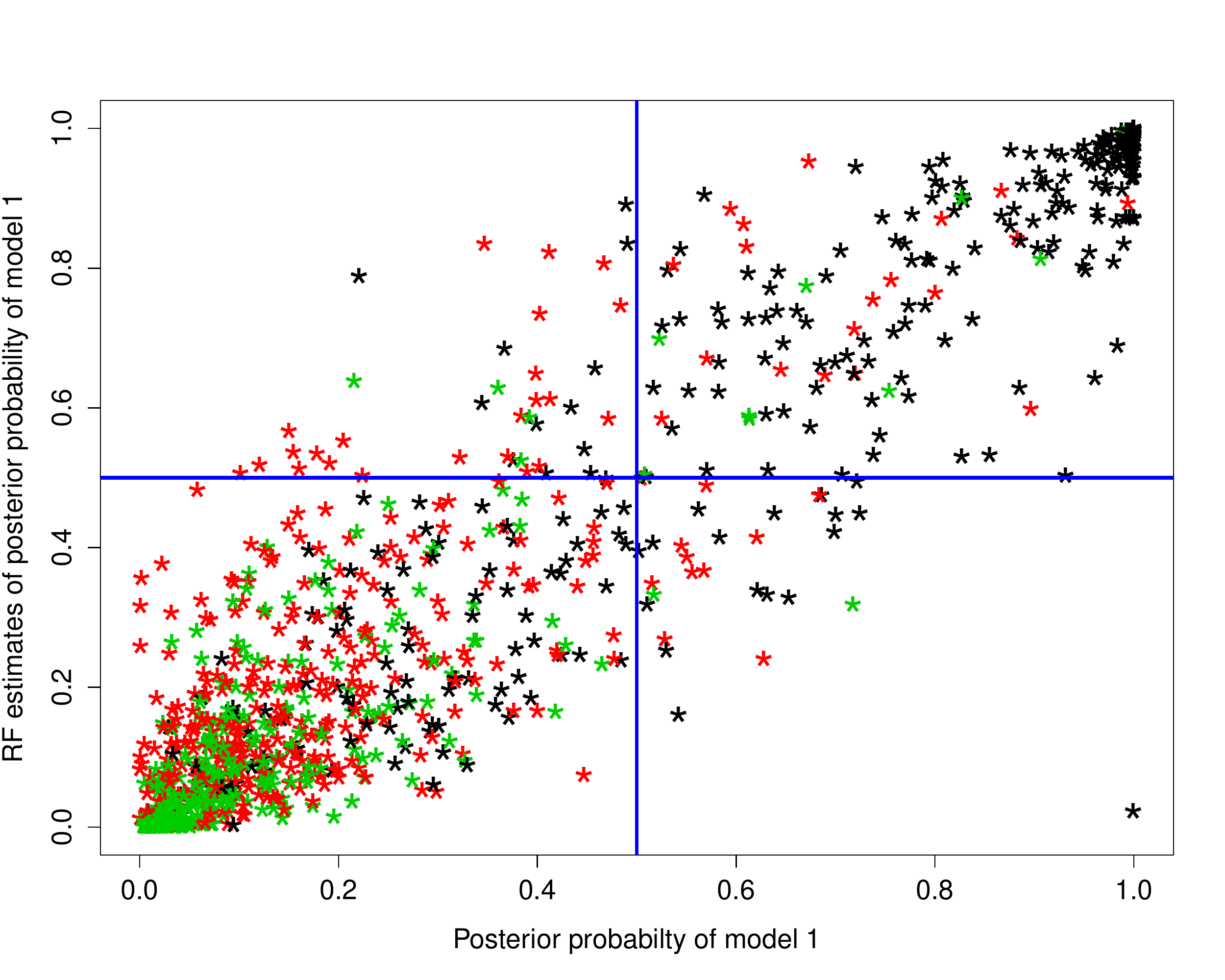}
\caption{\label{fig:toy_RFpost} True posterior probabilities of Model 1 against their Random Forest estimates 
for the test sample, with the same colour code as in Figure \ref{fig:toy_post}.}
\end{figure}

By comparison, the RF ABC model choice technique of \cite{pudlo:etal:2016} based on 500 trees
achieves an error rate of 0.276 on the test dataset. For this example, adding the two
LDA components to the summary statistics does not make a difference. This alternative procedure achieves
similarly good results in terms of prior error rate, since 0.276 is relatively closed to the absolute lower bound of 0.245.
However, as explained in previous sections and illustrated on Figure \ref{fig:toy_RFpost}, the RF estimates of the posterior
probabilities are not to be trusted. In short, a classification tool is not necessarily appropriate for regression
goals.

A noteworthy feature of the RF technique is its ability to be robust against non-discriminant variates. This obviously
is of considerable appeal in ABC model choice since the selection of summary statistics is an unsolved challenge. To
illustrate this point, we added to the original set of three summary statistics
variables that are pure noise, being produced by independent simulations from standard Gaussian distributions.
Table \ref{tab:RFnoise} shows that the additional error due to those irrelevant variates grows much more slowly than for
the standard ABC model choice technique, as shown in Table \ref{tab:ABCnoise}. In the latter case, a few extraneous
variates suffice to propel the error rate above 50 \%.

\begin{table}
\begin{tabular}{|l|l|}
Extra variables & prior error rate \\
\hline 0 & 0.276 \\
2 & 0.283 \\
4 & 0.288 \\
6 & 0.272 \\
8 & 0.280 \\
10 & 0.286 \\
20 & 0.318 \\
50 & 0.355 \\
100 & 0.391 \\
200 & 0.419 \\
1000 & 0.456 \\
\end{tabular}
\caption{\label{tab:RFnoise}
Evolution of the prior error rate for the RF ABC model choice procedure
as a function of the number of white noise variates.}
\end{table}

\begin{table}
\begin{tabular}{|l|l|l|}
Extra variables & optimal $k$ & prior error rate \\
\hline 0    & 20   & 0.277 \\
2    & 20   & 0.368 \\
4    & 140  & 0.468 \\      
6    & 200  & 0.491 \\
8    & 260  & 0.492 \\ 
10   & 260  & 0.526 \\
20   & 260  & 0.542 \\
50   & 260  & 0.548 \\
100  & 500  & 0.559 \\
200  & 500  & 0.572 \\
1000 & 1000 & 0.594 \\
\end{tabular}
\caption{\label{tab:ABCnoise}
Evolution of the prior error rate for a standard ABC model choice 
as a function of the number of white noise variates.}
\end{table}

\section{Human population genetics example}

We consider here the massive Single Nucleotide Polymorphism (SNP) dataset already studied in \cite{pudlo:etal:2016},
associated with a MRCA population genetic model corresponding to Kingman's coalescent that has been at the core of ABC
implementations from their beginning \citep{tavare:balding:griffith:donnelly:1997}. The dataset corresponds to
individuals originating from four Human populations, with 30 individuals per population. The freely accessible
public 1000 Genome databases {\sf +http://www.1000genomes.org/data} has been used to produce this dataset. As
detailed in \cite{pudlo:etal:2016} one of the appeals of using SNP data from the 1000 Genomes Project
\citep{genome:project:2012} is that such data does not suffer from any ascertainment bias.

The four Human populations in this study included the Yoruba
population (Nigeria) as representative of Africa, the Han Chinese
population (China) as representative of East Asia (encoded CHB), the British population (England and
Scotland) as representative of Europe (encoded GBR), and the population of Americans of African
ancestry in SW USA (encoded ASW). After applying some selection criteria described in \cite{pudlo:etal:2016},
the dataset includes 51,250 SNP loci scattered over the 22 autosomes with a median distance between two consecutive SNPs
equal to 7 kb. Among those, 50,000 were randomly chosen for evaluating the proposed RF ABC model choice method.

\begin{figure}
\centering
\includegraphics[width=0.4\textwidth]{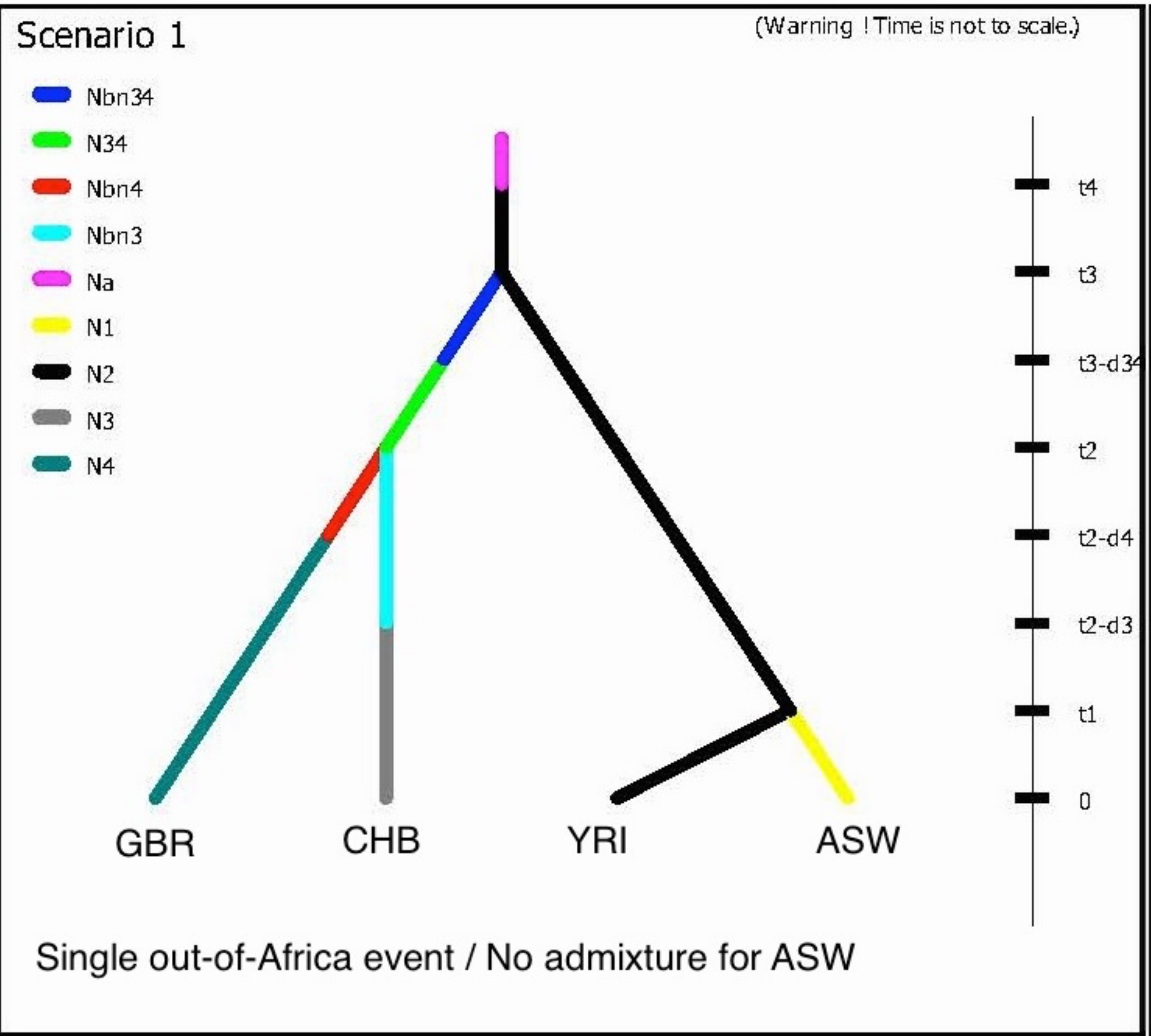}
\includegraphics[width=0.4\textwidth]{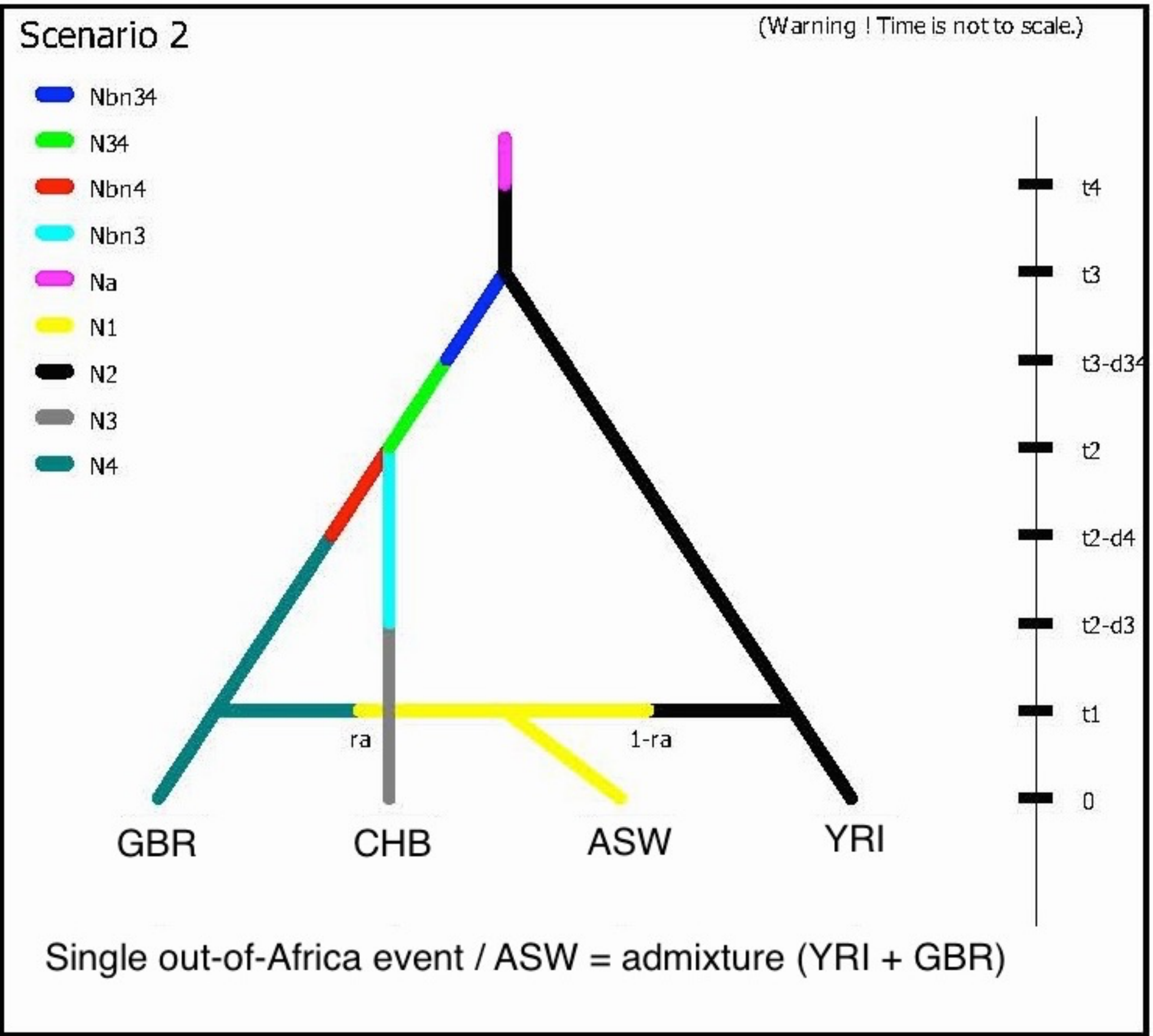}
\caption{\label{fig:scenarios} Two scenarios of evolution of four Human populations genotyped at 50,000 SNPs.
The genotyped populations are YRI = Yoruba (Nigeria, Africa), CHB = Han (China, East Asia), GBR = British
(England and Scotland, Europe), and ASW = Americans of African ancestry (SW USA).}
\end{figure}

In the novel study described here, we only consider two scenarios of evolution.  These two models differ by the possibility or
impossibility of a recent genetic admixture of Americans of African ancestry in SW USA between their African forebears and
individuals of European origins, as described in Figure \ref{fig:scenarios}.  
Model 2 thus includes a single out-of-Africa colonisation event giving an ancestral out-of-Africa population with a
secondarily split into one European and one East Asian population lineage and a recent genetic admixture of Americans of
African origin with their African ancestors and European individuals. RF ABC model choice is used to discriminate among
both models and returns error rates. The vector of summary statistics is the entire collection provided by the DIYABC
software for SNP markers \citep{cornuet:etal:2014}, made of 112 summary statistics described in the
manual of DIYABC.

Model 1 involves 16 parameters while Model 2 has an extra parameter, the admixture rate $r_a$.
All times and durations in the model are expressed in number of generations. The stable effective populations
sizes are expressed in number of diploid individuals.  The prior distributions on the parameters appearing in one of the
two models and used to generate SNP datasets are as follows:
\begin{enumerate}
\item split or admixture time $t_1$, $\mathcal{U}[1,30]$,
\item split times $\left(t_2,t_3,t_4\right)$, uniform on their support\\
$\left\{\left(t_2,t_3,t_4\right)\in[100,10000]^{\otimes 3}|t_2<t_3<t_4\right\}$, 
\item admixture rate (proportion of genes with a non-African origin in Model 2) $r_a\sim\sim\mathcal{U}[0.05,0.95]$, 
\item effective population sizes $N_1$, $N_2$, $N_3$, $N_3$ and $N_{34}$, $\mathcal{U}[1000,100000]$,
\item bottleneck durations $d_3$, $d_4$ and $d_{34}$, $\mathcal{U}[5,500]$,
\item bottleneck effective population sizes $Nbn_3$, $Nbn_4$ and $Nbn_{34}$, $\mathcal{U}[5,500]$,
\item ancestral effective population size $N_a$, $\mathcal{U}[100,10000]$,
\end{enumerate}

\begin{figure}
\centering
\includegraphics[width=0.8\textwidth]{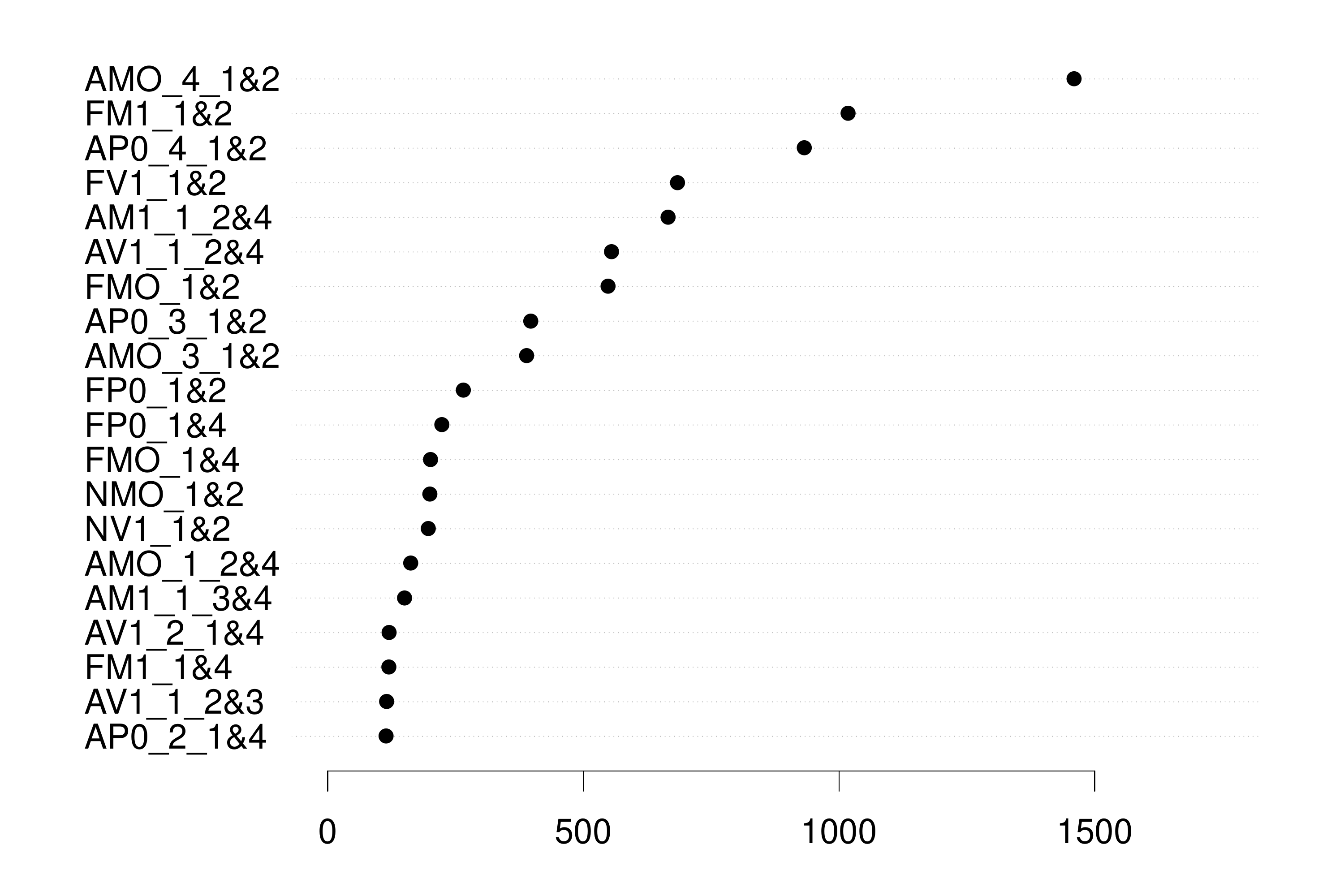}
\includegraphics[width=0.8\textwidth]{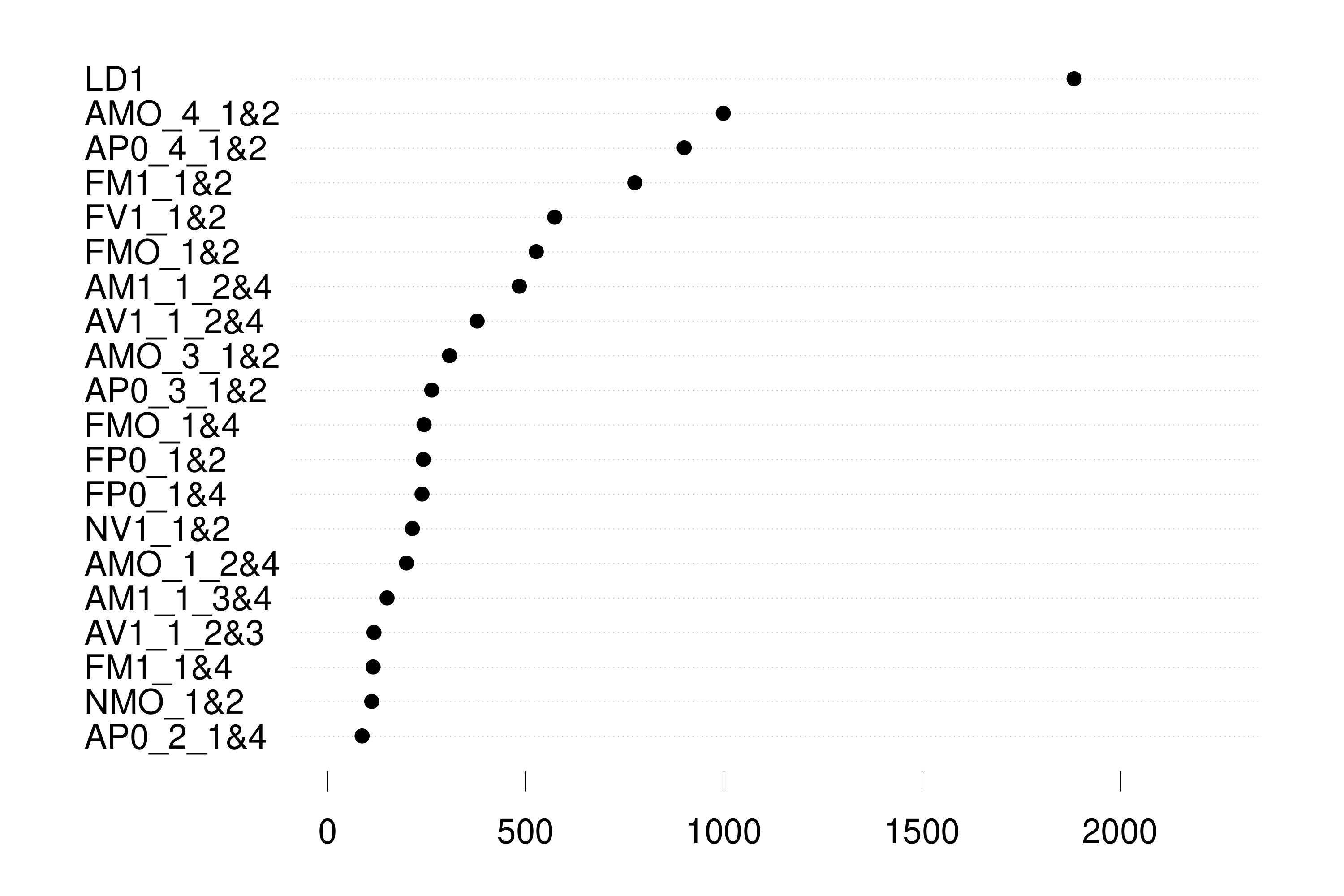}
\caption{\label{fig:human_VI} Contributions of the most frequent statistics in the RF.
The contribution of a summary statistic is evaluated as the average decrease in node impurity at all nodes where it
is selected, over the trees of the RF when using the 112 summary statistics {\em (top)} and when further adding the
first LDA axis {\em (bottom)}.}
\end{figure}

\begin{figure}
\centering
\includegraphics[width=0.8\textwidth]{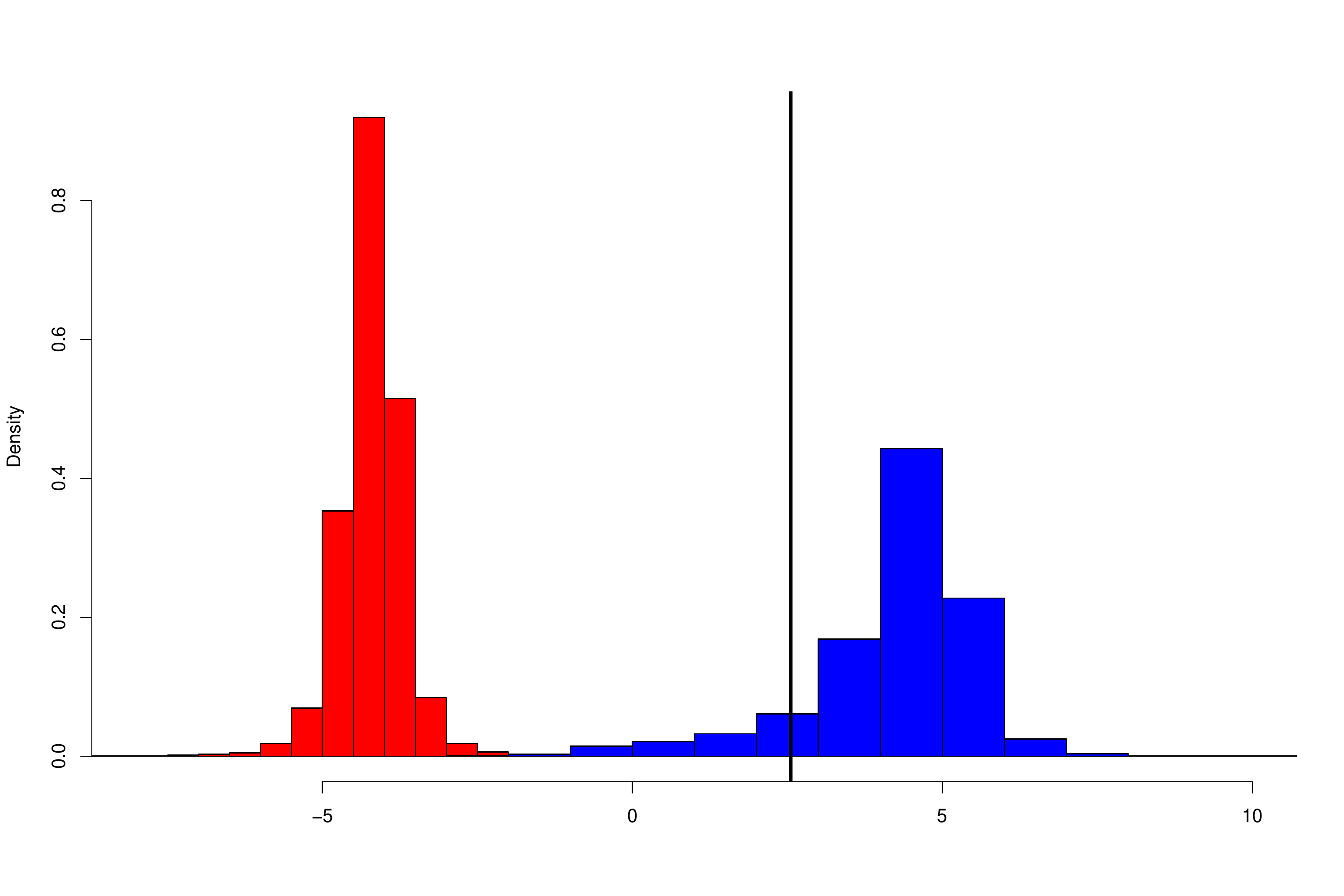}
\caption{\label{fig:human_LDA} Distribution of the first LDA axis derived from the reference table, in red for Model 1 and in blue for Model 2.
The observed dataset is indicated by a black vertical line.}
\end{figure}

For the analyses we use a reference table containing 19995 simulations: 10032 from Model 1 and
9963 from Model 2. Figure \ref{fig:human_LDA} shows the distributions of the first LDA projection
for both models, as a byproduct of the
simulated reference table. Unsurprisingly, this LDA component has a massive impact on the RF ABC model choice procedure.
When including the LDA statistic, most trees (473 out of 500) allocate the observed dataset to Model 2.
The second random forest to evaluate the local selection error leads a high
confidence level: the estimated posterior probability of Model 2 is greater than $0.999$. 
Figure \ref{fig:human_VI} shows contributions for the most relevant statistics 
in the forest, stressing once again the primary role of the first LDA axis. Note that using solely this first LDA axis increases
considerably the prior error rate.

\section{Conclusion}

This chapter has presented a solution for conducting ABC model choice and testing that differs from the usual practice
in applied fields like population genetics, where the use of Algorithm \ref{algo:ABCMoo} remains the norm. This choice
is not due to any desire to promote our own work, but proceeds from a genuine belief that the figures returned by this
algorithm cannot be trusted as approximating the actual posterior probabilities of the model. This belief is based on
our experience along the years we worked on this problem, as illustrated by the evolution in our papers on the topic.

To move to a machine-learning tool like random forests somehow represents a paradigm shift for the ABC community. For
one thing, to gather intuition about the intrinsic nature of this tool and to relate it to ABC schemes is certainly
anything but straightforward. For instance, a natural perception of this classification methodology is to take it as a
natural selection tool that could lead to a reduced subset of significant statistics, with the side appeal of providing a natural
distance between two vectors of summary statistics through the tree discrepancies. However, as we observed through
experiments, subsequent ABC model choice steps based on the selected summaries are detrimental to the quality of
the classification once a model is selected by the random forest. The statistical appeal of a random forest is on the
opposite that it is quite robust to the inclusion of poorly informative or irrelevant summary statistics and on the
opposite able to catch minute amounts of additional information produced by such additions. 

While the current state-of-the-art remains silent about acceptable approximations of the true posterior probability of a
model, in the sense of being conditional to the raw data, we are nonetheless making progress towards the production of
an approximation conditional on an arbitrary set of summary statistics, which should offer strong similarities with the
above. That this step can be achieved at no significant extra-cost is encouraging for the future. 

Another important inferential issue pertaining ABC model choice is to test a large collection of models. The
difficulties to learn how to discriminate between models certainly increase when the number of likelihoods in
competition gets larger. Even the most up-to-date machine learning algorithms will loose their efficiency if
one keeps constant the number of iid draws from each model, without mentioning that the time complexity will
increase linearly with the size of the collection to produce the reference table that trains the
classifier. Thus this problem remains largely open.

\end{document}